# Evaluation of Distributed Intelligence on the Smart Card

**Kazuo J. Ezawa**     **Greg Napiorkowski**     **Mariusz Kossarski**
Mondex International Limited
Atlantic Technology Center
Suite 109,100 Campus Drive,
P.O. Box 972, Florham Park,
New Jersey, 07932-0972, USA

**Abstract**

We describe challenges in the risk management of smart card based electronic cash industry and describe a method to evaluate the effectiveness of distributed intelligence on the smart card. More specifically, we discuss the evaluation of distributed intelligence function called "on-chip risk management" of the smart card for the global electronic cash payment application using micro dynamic simulation. Handling of uncertainty related to future economic environment, various potential counterfeit attack scenarios, requires simulation of such environment to evaluate on-chip performance. Creation of realistic simulation of electronic cash economy, transaction environment, consumers, merchants, banks are challenge themselves. In addition, we shows examples of detection capability of off-chip, host based counterfeit detection systems based on the micro dynamic simulation model generated data set.

# 1 INTRODUCTION

The smart card market is expanding rapidly as a result of its superior security, reliability, and capacity. Its ability to carry intelligent applications on the card such as "access", "credit/debit", "electronic cash", etc. gives the smart card an expanding market. The smart card provides distributed processing power, a computer in your wallet.

Smart card has effective card authentication and verification methodologies, employing cryptographic techniques. Smart card can be authenticated in one of two ways either Static or Dynamic:
1. Using Static authentication the smart card sends the terminal a "digital signature" containing information which uniquely identifies the card e.g., card serial number, manufacture ID and manufacture date. The terminal will decrypt the signature to determine if the card is genuine
2. Using Dynamic authentication the terminal will generate some random data, known as seed, and will ask the smartcard to encrypt the data. On receipt of the encrypted data the terminal will decrypt the data. If the decrypted data is the same as the seed then the card is genuine. Dynamic authentication is only possible with smartcards due to their ability to perform cryptography.

As card industries move from magnetic strip cards to smart cards, ability to process information on the cards drastically increases. In the case of magnetic strip card, it is imperative to rely on the host system's intelligence to authorize the transactions (e.g., credit/debit) since it has no information processing capability of its own. As we move to smart card, the intelligence doesn't have to be concentrated on the host system, but it can be moved from the host system to more balanced combination of host and smart card itself.

## 1.1 DISTRIBUTED INTELLIGENCE ON SMART CARD AS RISK MANAGEMENT TOOL

Security and risk management are integral parts of development and deployment of "risk managed" smart card application for a global electronic cash payment such as Mondex electronic cash. There are three critical components, -- prevention, detection, and containment, -- to achieve balanced risk managed smart card application. The security is primarily concerned with "prevention." The risk management is primarily concerned with "detection" and "containment" in the event that the security were to be broken. The discussion of security can be found in [Maher, 1997].

The objectives of smart card electronic cash risk management can be summarized as follows:

- To contain the economic risk exposure to a predetermined level, and
- To ensure the stability and continuity of the product.

One of the key economic risk exposures is due to "counterfeit" of electronic currency. Among other things, the security and risk management is designed to address this threat head-on to minimize the impact of such attacks. At the same time, it is designed to ensure the stability and continuity of the product.

More specifically, to accomplish smart cart electronic cash risk management objectives, risk management strategy can stand on the four pillars:
- Prudential Risk Management
- On-Chip Risk Management
- Off-Chip Risk Management
- Micro Dynamic Simulation

Each pillar has its unique contribution to the objectives, but when they are balanced and combined, they become a formidable structure to base the risk management strategy, and to accomplish the objectives. It may seem obvious, but the prudential risk management is essential to the success of the product. It includes corporate governance and structural control. It is the foundation for the rest of the risk management is build onto.

One of the fundamental strategies in smart card electronic cash risk management such as Mondex is to economically exploit the on-chip data processing power of the smart card to the maximum extent. By installing risk management functionality on a chip, some of the critical risk management tasks are performed at the time of transaction autonomously on the transacting smart cards. On-chip risk management functionality includes both **on-chip detection**, and **on-chip incidence response**. On-chip incidence response can be activated autonomously, or by the central command.

There's a paradigm shift in **off-chip** (i.e., host system based) risk management as well. It partly relies on the on-chip intelligence to collect information selectively. At the same time, a multi-layered off-chip monitoring and detection capability is deployed to analyze possible counterfeit activities. All the on-line transactions can be monitored, and some of the off-line transactions are selectively monitored.

Since counterfeit activities on electronic cash purses/cards are non-existent, **Micro Dynamic Simulator** was developed to simulate the impact of various counterfeit scenarios on the electronic cash economy for Mondex. It allows us to evaluate the effectiveness of the on-chip detection, the on-chip incidence response, and off-chip detection systems. It also generates data sets to create off-chip detection models. As we succeed in risk management, counterfeit transactions won't be available. The evaluation of new enhancement to on-chip functionality and the re-calibration of off-chip detection models have to come from simulator using real market inputs.

The paper is organized as follows. Section 2 describes Mondex global electronic cash payment scheme to set the stage. Section 3 discusses the distributed intelligence – on-chip risk management capability on the smart card as an example of such intelligence. Section 4 discusses the micro dynamic simulation. Section 5 discusses the quantification of impact of counterfeiter's threat scenarios using micro dynamic simulator. Section 6 discusses the effectiveness of off-chip, host system based counterfeit detection systems. Section 7 summarizes the discussion.

## 2 GLOBAL SMART CARD BASED ELECTRONIC CASH PRODUCT

The global smart card based electronic cash product such as Mondex electronic cash has the security and the risk management to prevent, detect, contain, and recover from potential counterfeit activities. It is designed to make counterfeiter's "chain" of tasks as difficult as possible in every step of the way [Ezawa *et al.* 1998].

The product is designed for the efficient electronic cash payment transactions. It performs purse (chip) to purse (chip) transactions without central authorization. It has many on-chip capability and features such as physical security, cryptographical security, purse class structure (i.e., it restrict the interactions of different type of purses), purse limit, on-chip risk management capability (e.g., credit turnover limit), and migration[1]. Purse class structure, purse limit, credit turnover limit will be revisited in the following section.

Figure 1 shows the Mondex transactions among the different classes of purses. Solid line indicates transactions currently allowed, and dotted line indicates the transactions severely restricted (or disallowed) at this stage of product evolution.

Ideally, an advanced smart card based electronic cash scheme, as a substitute for "real" money, should parallel the existing money supply and banking system.

---

[1] It involves switching of one public key scheme to the other.

Therefore such a scheme would include a currency "originator" (equivalent of central bank), and "members" (commercial banks and other financial institutions with their branches). There are merchants who transact with consumers and members, and consumers transacting with other consumers, merchants, and members.

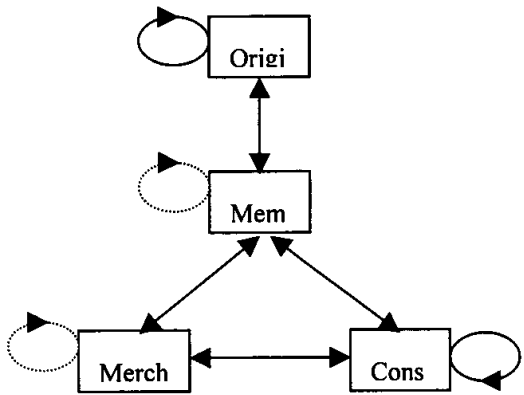

Figure 1: Transactions among different class of purses

In the following we discuss non-security related features:

Chip to Chip: The value (electronic cash) is transferred from payer purse (chip) to payee purse (chip) without third party authorization.

Purse Class Structure: It classifies purses into different types of purses and determines what types of purses can transact with each other. Each purse can transact only with predetermined list of purse classes. For example, a consumer purse which is linked to the purse holder's direct deposit account of the member can transact with other consumer purses, two types of member purses, and three types of merchant purses.

Purse limits: High value limit purses (such as originator, and member purses) are monitored on line. All transactions to and from these purses are recorded and monitored (e.g., merchants and consumer deposit transactions.) Consumer purses are expected to have relatively low purse limits (e.g., up to $1000.)

Credit turnover limit: An on-chip risk management capability to monitor amount of value being received by a consumer purse from non-member purses such as consumer to consumer, and merchant refund transactions. If the transaction causes the credit turnover limit to be exceeded, this on-chip logic suspends a part or whole features of the purse. The credit turnover limit is customized by members to fit for the purse holder's normal needs.

## 3 DISTRIBUTED INTELLIGENCE -- ON-CHIP RISK MANAGEMENT

As we have already discussed, one of the fundamental strategies in smart card electronic cash risk management such as Mondex electronic cash is to economically exploit the on-chip data processing power of the smart card to the maximum extent. It allows risk management tasks done on the chip autonomously for each transaction without external intervention.

On-chip functionality in the "security" arena has been around many years, but in the risk management arena it is a new and relatively unexplored field. In the past, an old generation of simple chips with a limited computing capability forced to rely heavily on the host systems for intelligence for transactions and monitoring (i.e., on-line transactions and authorization.) New generation of chips have more computing power and memory, allow them to have "distributed" intelligence (i.e., "intelligent agent") on the on-chip itself as opposed to "central" intelligence on the host systems somewhere in the network infrastructure.

It has advantages in the effectiveness and timeliness of on-chip risk management functionality. It allows real-time information gathering, monitoring & detection, and incidence response at the time of purse transactions. It also has advantage in the efficiency in data processing. It processes transaction data on-chip for risk management at the time of transaction as opposed to massively accumulated transaction data at host system a few days later

For example, if we have 100 million cards, a distributed intelligent agent makes decisions for a few transactions a day for each card at the time of each transaction, whereas a central intelligent agent makes decisions for a few hundred million transactions a day, a few days later. For the central agent to make a decision, it needs a database of a year or two worth of these transactions. It is truly requires substantial investment to perform this task in near real-time basis. For a micro payment scheme, it is difficult to see the cost/benefit justification when we have the alternative method.

The on-chip risk management capability is protected by the chip (tamper resistant). To disable its capability, it has to pass the layers of the security of the chip.

One of the critical elements and advantages of the on-chip risk management capability is that it continuously functions even under complete physical security breakdown. Yes, it is true that the risk management functionality of the compromised chip will be disabled. But for the counterfeiters to benefit from their activities, i.e., to obtain economic gain, they need to interact with other legitimate purses (cards) which still have active and functioning on-chip risk management capability which are unique to each purse. Its wide range of functionality is discussed in the next subsection, but it is a formidable tasks to pass all the screens without triggering some actions on the on-chip risk management part.

Lastly, it is more cost effective to invest in on-chip risk management functionality than that of off-chip (host) risk management infrastructure to perform the same functionality. Although risk management must invests in the off-chip (host based) risk management detection, not to duplicate the functionality of on-chip, but to complement. Each has unique capability to contribute to the overall risk management. Off-chip risk management discussion can be found in [Ezawa, et. al. 1999].

### 3.1 DISTRIBUTED INTELLIGENCE PORTFOLIO – ON-CHIP RISK MANAGEMENT PORTFOLIO

There are two primary methods for fraud and counterfeit detection in general, one measures the "velocity" of transactions, and the other compares transactions against "statistical signature" of the purse. It is true for both on-chip and off-chip (i.e., host system based) detection. The "velocity" method, which monitors amount and volume of transactions, is widely used in the telecommunications, and financial industries to monitor potential fraudulent transactions. The "statistical signature" method, which monitors transactions against the past behavioral patterns, is more computationally intensive and requires more infrastructure support. It is also widely used in the telecommunications and financial industries to monitor net bad debt as well as fraudulent transactions and accounts [Ezawa, 95 & 96].

Risk management can use both "velocity" and "statistical signature" methods in on-chip as well as off-chip risk management. The "credit turnover limit" is a good example of "velocity" based method implemented as the on-chip risk management monitoring and detection capability.

On-chip risk management does not rely on single functionality but a portfolio of on-chip functionality to meet the various potential counterfeit threats. There are five complementary components to the on-chip portfolio[2]:
- On-Chip Information Gathering (for Detection)
- On-chip Control
- Ability to Target Purses (for On-Chip Response)
- On-Chip Response
- Security

On-chip functionality allows risk management tasks done on the chip autonomously for each transaction without external intervention. The on-chip risk management capability performs the following two functions extremely well:
- Early Warning of Counterfeit Activities
- Containment of Counterfeit Activities at the point of transaction

<u>Early Warning</u>: It provides earliest possible warning of the counterfeit activities. It provides such a warning even if the counterfeit activity is low (in term of volume.) On-chip logic can provide warning at the initial stage of counterfeit activities, and allow incidence response team to react to the counterfeit threat, and contain it early. The containment task will be easier at the early stage and less costly.

<u>Containment</u>: By constraining the functionality of fraudulent purses, it contains/restrains the flow of counterfeit values to the market. Fraudulent population is defined as the population who is knowingly and willingly buys counterfeit value at discount from counterfeiters and their collaborators. When counterfeiters and their collaborators try to use/load the purses to distribute counterfeit values, most of these fraudulent purses are quickly disabled.

### 3.2 ON-CHIP INCIDENCE RESPONSE

As we discussed, the on-chip risk management has on-chip incident response capability in an autonomous mode. Alternative approach to the counterfeit contingency is at the chip level by a central command to activate on-chip incidence response on a contaminated segment of purses (cards.) Once it activated, it will function autonomously without outside intervention (i.e., host systems). It is the fastest way to respond to the potential incident.

The central command & communication (C3) mechanism is based on chip to chip secure communication. It has two aspects to it. One is the transmission and exchange of messages, and the other is the conduit and storage of information on the smart card. The primary task of C3 is to change the parameters on the purse, on-chip risk management, and security renewal remotely and securely by sending an authenticated system message.

---

[2] Note that in Mondex a portfolio of on-chip functionality is in various stages of development, some are implemented and deployed, and others are under evaluation and planning. It will follow a natural product cycle of continual renewal of risk management capability to meet the future challenges.

### 3.3 CENTRAL COMMAND & COMMUNICATION (C3):

1. Provides ability to send authenticated system message transmission to other contacting purses
2. Allow dynamic re-customization of on-chip risk management parameters if required when the purse contact with other purses (e.g., consumer purses)
3. Facilitate the targeting on-chip response to specified segment of purses
4. Provide ability to reset on-chip parameters (e.g., by member purses)
5. Facilitate loading / unloading of purse related pay loads (purse upgrades, security renewal, etc.)

## 4 MICRO DYNAMIC SIMULATOR

To quantify a threat scenario, one needs to observe or model, the following phases: 1) Creation of counterfeit value, 2) Interaction of electronic purses (transactions), 3) Diffusion of both legitimate and counterfeit value throughout the economy, and 4) Incident responses (countermeasures).

Micro Dynamic Simulator (simulator for short) is a particular application of the micro dynamic simulation concept to the electronic cash scheme. The model's design is flexible enough to reflect not just today's but also other possible future scheme structures. The simulator was used to assess the effectiveness of the selected responses against the likely threats.

The task to quantify a threat scenario requires, among other information, data on individual purses' transactions as well as on the effectiveness of the on-chip based response. Therefore we use the *micro dynamic simulation model*. In general, it is a computer model that imitates the dynamics of the electronic cash scheme. It has the following important features: 1) Mimics the expected longer term evolution of the electronic cash scheme, 2) Reflects, through respective model parameters, short term behavioral patterns, e.g. seasonal fluctuations, 3) Follows the transaction behavior of individual purses, e.g. a number and frequency of transactions, and 4) Keeps a complete record of all individual transactions. For example, following characteristics are simulated:

- Three time segments (primary period (e.g., month), secondary (e.g., day), and tertiary-period (e.g., hour))
  - Four types of transactions (deposit, withdrawal, purchase, consumer to consumer)
  - Active card normally distributed
  - Number of transaction by Poisson distribution
  - Amount in general Normally distributed
- On-chip risk management on each purse.
- Birth/Death and population growth of originators, members, merchants, and consumers (and associated purses)
- Consumer and merchant circles
  - Number of Consumer & Merchants Normally distributed
  - In-Circles probability Assignment

The features described above allow an analyst to perform various experiments. The essence of every experiment is to: 1) Design a threat scenario and inject the related counterfeit value into the system, and 2) Build in and invoke during the simulation the on-chip and off-chip based responses.

The attached appendix shows examples of input and output screens of the simulator. To increase model's flexibility and the level of detail, as far as the transaction patterns are concerned, each level of scheme participants can be further segmented. Segments within the same level of participants differ from each other by their respective transaction patterns, as defined, for instance, by number and type of daily transactions.

Figure 2 shows the main screen of Mondex Micro Dynamic Simulator. In this simulation model, there are 2 member bank segments (one bank segments and one counterfeit bank segment). There are three consumer segments (consumer, fraudulent consumer, and counterfeiter). There are two merchant segments. "Simulator" node allows us to define the simulation property, such as duration, starting date, etc. "MXICA" represents "Certificate Authority" node, and allows us to send C3 commands, "value creation", etc. "Originator" node controls the circulation of the currency in the simulated territory (e.g., country).

Figure 3 in the Appendix shows a window that defines a member segment given the originator. It allows the user to specify various characteristics of the member segment, ranging from, for example, member type (merchant bank, consumer bank, or both) to birth/death rates for members, merchants and consumers (i.e.

population growth and decline.) Member segments can be declared as counterfeit segments by clicking the corresponding "counterfeit" check box. Note that, at the purse level, the simulator keeps tracks of individual purse setting such as purse limit, value balance and on-chip risk management functionality.

An ability to produce and analyze multiple runs of the simulator model under different scenarios allows the user to experience the management of the electronic cash economy before the scheme is actually rolled out.

The risk management capabilities need to be continuously upgraded to match new potential threats in the rapidly evolving electronic commerce. The simulator model plays a critical role in the evaluation of both on-chip and off-chip new risk management tools to anticipate and prepare for the future counterfeit challenges.

In addition to being a tool to evaluate the impact of counterfeit scenarios, the simulator model also generates transactions that can be used to train off-chip detection model(s). The simulator model is to be calibrated for every respective currency originator (i.e. country) to reflect the particular behavior of its purse users and their transaction patterns of their territories.

The simulated diffusion of the counterfeit value and an effectiveness with which it can be detected and contained provide the critical information that allows us to quantify a threat scenario in question.

We briefly described the micro dynamic simulator under development. This simulator can provide quantitative information to analyze the effectiveness of on and off chip risk management schemes. It will be also useful for recruiting new members, satisfying financial authorities, as well as existing members by demonstrating and quantifying the security and risk management issues.

## 5  EVALUATION

We evaluated the above mentioned detection systems in the "Street Corner Counterfeit Value Distribution Threat Scenario" that is discussed in [Ezawa, *et al.*, 1998]. This is still a preliminary result. This counterfeit threat scenario assumes that the counterfeiters will sell, at a discount, counterfeit electronic cash to a fraudulent population, in exchange for "real" local currency. The fraudulent population is defined as the one that would engage in such transactions knowingly and willingly. The fraudulent population is not necessarily as loyal as agents of counterfeit organization and the "secret" is bound to be leaked to the law enforcement institutions or electronic cash issuing institution.

It showed that this is quite a difficult task to carry out flawlessly. For the sake of the evaluation of on-chip risk management capability, we assumed the following:
- Counterfeit organization has a well financed, well established world wide network, and a large number of dedicated agents in place.
- It successfully broke the security of the chip / purse application on the smart card that required a complete secrecy over an extended period of time while various tasks are performed to break security.
- It created a counterfeit electronic cash application -- "shrink wrap" product of "golden goose" that can generate counterfeit electronic cash with flawless imitation of electronic cash application (e.g., Mondex purse) functionality.
- It established counterfeit value distribution channels with no "informants".
- Counterfeiter/agents can correctly identify "fraudulent" population who is willing to buy counterfeit values with discount. They never make mistakes. If they approach a normal/honest person, he or she might inform the financial institution or authority.

### 5.1  COUNTERFEIT ATTACK SCENARIO

Simulation model was set to run 180 days and the counterfeit attack starts at the last 6 days. The length of the run is set so that simulation transaction data will provide significant amount of normal transactions. One the first day of the attack, April 1, 1998, the counterfeiters inject a very small amount of counterfeit value to the electronic cash economy to test the system. On the second day, April 2, 1998, they inject amount they desire. On the third day, April 3, they stop their activities completely to observe and evaluate their performance of the previous day. They resume the counterfeit value distribution for the rest of the three days. Note that the calendar days are important, since the simulator simulates the day of the week, the seasonality and holiday impacts to the behaviors of various consumer and merchant segments.

### 5.2  RESULTS

In this section, we show the effectiveness of credit turnover limit only. Due to security reasons, although the central command based security renewal and dynamic re-customization are found to be very effective, the discussion is omitted.

### 5.2.1 Automatic Response – Credit Turnover Limit

Figure 4 in the Appendix shows the impact that counterfeit activities have on the number of locked up purses. This is the direct effect of the on-chip risk management functionality. The locked up purses are the legitimate ones used by fraudulent population that happen to be contacted by the counterfeit purses in order to receive the created counterfeit value. When a preset condition is met, the on-chip risk management functionality turns on on-chip response autonomously in this case locking up the purses. It turned out that almost all the fraudulent purses were locked up due to the credit turnover limit and forced them to visit member (bank) for re-customization (resetting of on-chip logic). Member can retrieve the information on the locked purse and can determine that possible counterfeit activities are in present.

## 6 OFF-CHIP, HOST SYSTEM BASED COUNTERFEIT DETECTION

Off-chip, host system based counterfeit detection systems complements the detection of counterfeit activities based on on-chip risk management capability. There are four monitoring systems in the Mondex electronic cash scheme, two (i.e., currency and member monitoring systems) reside in the originator monitoring over a country or a territory, and another two (i.e., merchant and consumer monitoring systems) reside in the issuing members (banks). In the following we show sample performance of two (currency and merchant monitoring systems) based on this attack scenario.

### 6.1 CURRENCY MONITORING SYSTEM

The objective of currency monitoring system is to detect the presence of potential counterfeit value in (almost) real time for the three types of attacks; rapid (i.e. a sudden redemption of counterfeit value), moderate, and long term (skimming). And to provide recommendations as to what steps should be taken to identify sources for the potential counterfeit value, once detected.

The methodology of detecting the potential counterfeit value rests on the fact that any injection of counterfeit value into the Mondex economy will be eventually deposited with the originator and redeemed for "regular" money. Consequently, an unusual surge in the redeemed electronic cash value should be carefully scrutinized.

Figure 5 shows the detection of counterfeit injection based on rolling monthly statistical model. As shown, it detects on the second day of the attack. In general, rolling monthly model can detect smaller counterfeit attack than that of daily or rolling weekly model. It of course takes into account the weekly, and monthly seasonality of the currency behavior.

### 6.2 MERCHANT MONITORING SYSTEM

The objective of the Merchant Monitoring System is to detect the presence of the counterfeit value in a timely fashion and provide decision support when the potential incursion of the counterfeit value is detected.

Merchant Monitoring System primarily tracks the electronic cash value transactions between individual merchant and its acquiring member. These transactions, i.e., value transfer from a merchant to a member are fully accounted. One can also conclude that the merchant transaction values should be transformed in order to properly apply various statistical tests and model estimation methods. The preliminary evidence suggests, for instance, that the nominal transaction values are approximately log-normally distributed.

Similarly to currency monitoring system, separate models are build to detect the three different types of attacks, rapid, moderate, and long term ("skimming") attacks that would be carried out via merchant purses while taking into account the length of a given merchant history.

Figure 6 shows the detection of counterfeit injection based on rolling weekly statistical model. As shown it detects the out of bound condition quickly. As it turns out, the merchant monitoring system is a very effective detection system because it monitors the historic redemption patterns of the merchant. Even relatively small amount of counterfeit value spending by the counterfeit or fraudulent population can impact the merchant redemption pattern sufficiently to alarm the merchant monitoring system. When there is a attack, it causes a large number of merchants to be in the out of norm condition (red stars) that normal, and it gives us warning of potential counterfeit activities.

Overall, both systems are very effective to detect material counterfeit attacks. Each system has its own strength and weakness, and the use of four different types of monitoring systems will compensate each other for respective weakness of the individual system.

## 7 SUMMARY

We discussed the risk management of smart card based electronic cash industry and a method to evaluate the effectiveness of distributed intelligence function called "on-chip risk management" of the smart card for the global electronic cash payment application using micro dynamic simulation. We found that it is critical to evaluate the distributed intelligent capability quantitatively using micro dynamic simulation. We demonstrated the effectiveness of distributed intelligence – credit turnover limit to be very effective in detecting and containing counterfeit activities. We showed examples of detection capability of off-chip, host based counterfeit detection systems based on the micro dynamic simulation model generated data set, and found to be very effective.

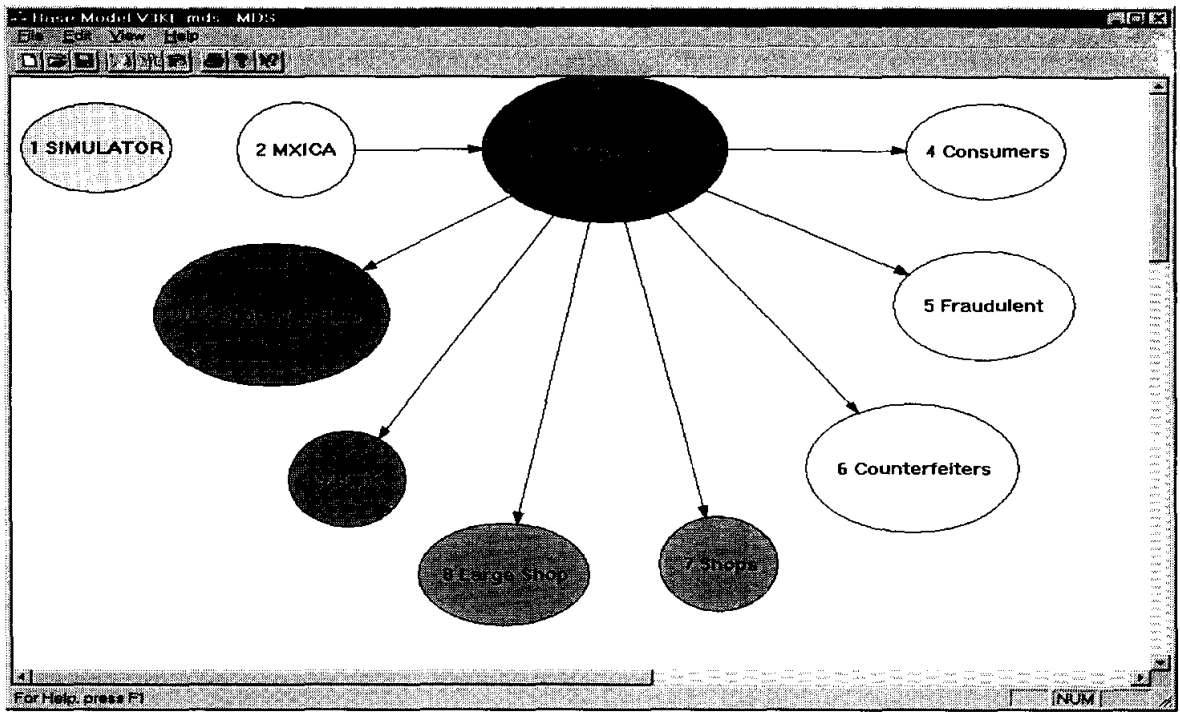

Figure 2: Example Main Screen

Figure 3: Example Input Screen - Member Segment Specification

Figure 4: Example Output – On-chip Response to Counterfeit Activity

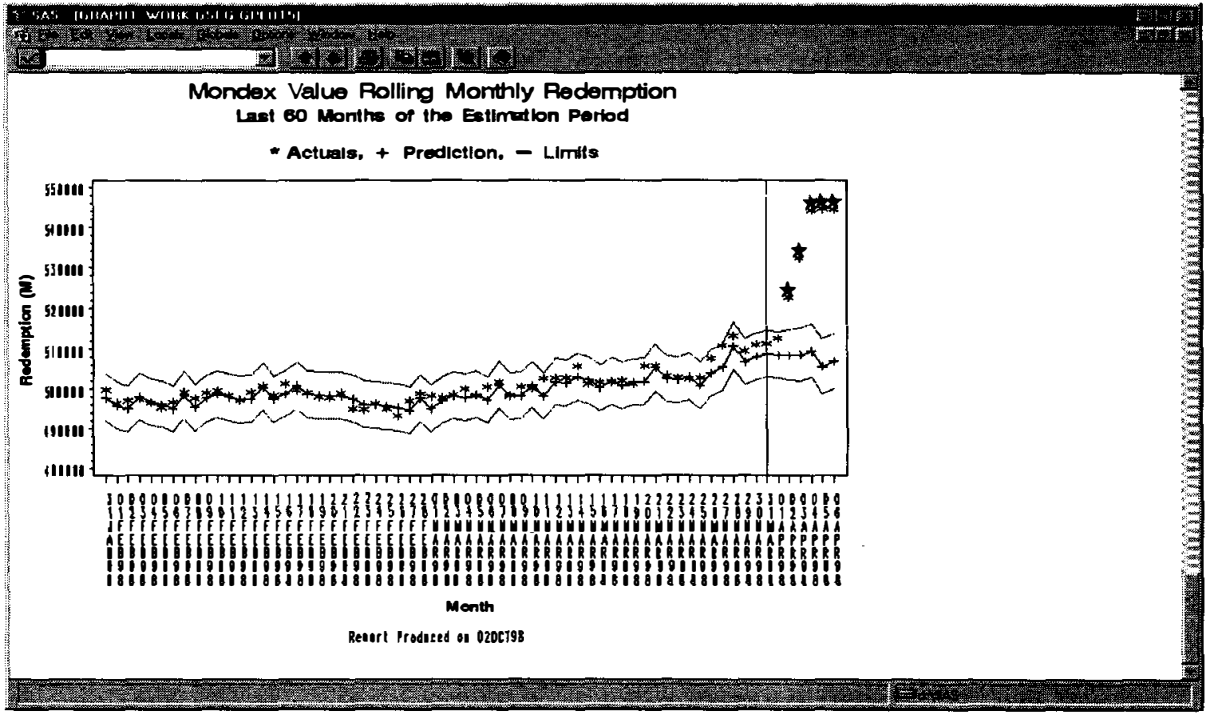

Figure 5: Originator Currency Monitoring System

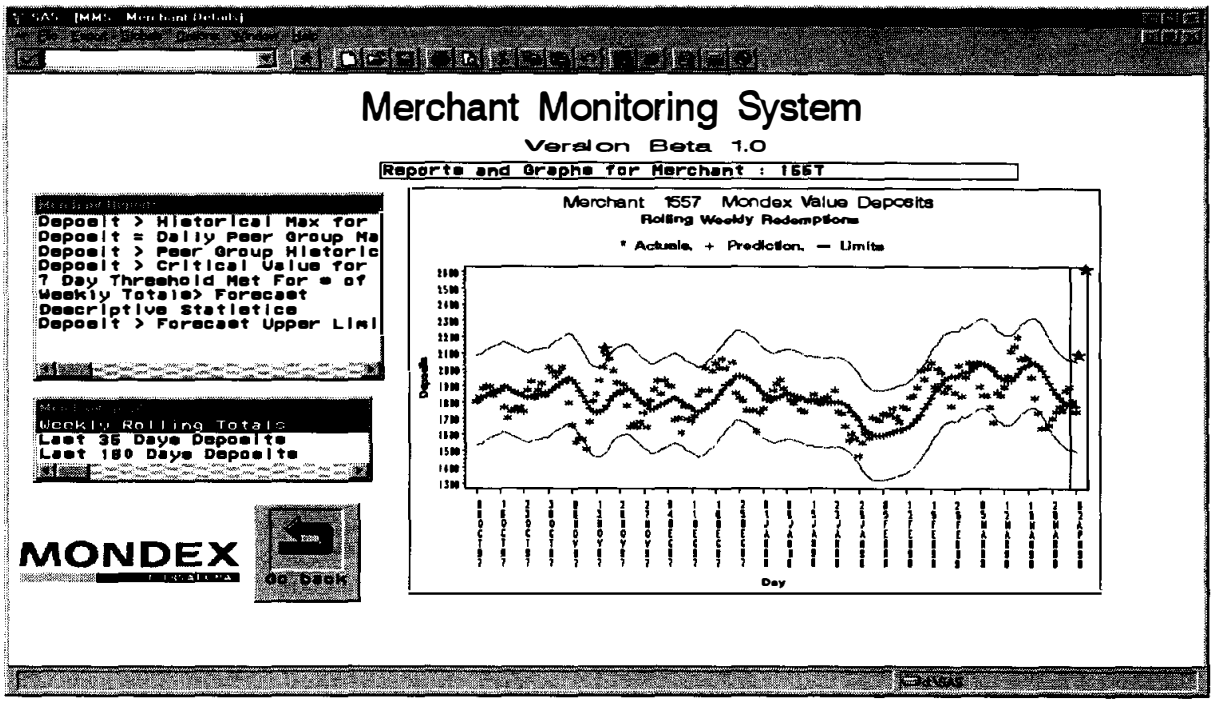

Figure 6: Merchant Monitoring System